\documentclass[prd,preprint,
superscriptaddress,tightenlines,nofootinbib, eqsecnum]{revtex4-2}

\usepackage{amsmath}
\usepackage{amsfonts}
\usepackage{amssymb}
\usepackage{bm}
\usepackage{hyperref}
\usepackage{mathrsfs}
\usepackage{graphicx}

\usepackage{empheq}
\usepackage{ulem}
\usepackage[usenames]{color}
\definecolor{darkgreen}{rgb}{0,0.5,0}

\hypersetup{
 bookmarks=true,         % show bookmarks bar?
 unicode=false,          % non-Latin characters in Acrobat???s bookmarks
 pdftoolbar=true,        % show Acrobat???s toolbar?
 pdfmenubar=true,        % show Acrobat???s menu?
 pdffitwindow=false,     % window fit to page when opened
 pdfstartview={FitH},    % fits the width of the page to the window
 pdftitle={My title},    % title
 pdfauthor={Author},     % author
 pdfsubject={Subject},   % subject of the document
 pdfcreator={Creator},   % creator of the document
 pdfproducer={Producer}, % producer of the document
 pdfkeywords={keyword1} {key2} {key3}, % list of keywords
 pdfnewwindow=true,      % links in new window
 colorlinks=true,       % false: boxed links; true: colored links
 linkcolor=red,          % color of internal links
 citecolor=cyan,        % color of links to bibliography
 filecolor=magenta,      % color of file links
 urlcolor=darkgreen,           % color of external links
 linktocpage=true
}

\DeclareSymbolFontAlphabet{\mathrsfs}{rsfs}
\DeclareMathAlphabet{\mathcal}{OMS}{cmsy}{m}{n}

\allowdisplaybreaks

\newcommand{\dd}{\mathrm{d}}

\begin{document}

\title{Absence of gravitational polarization mechanism \\in the canonical bimetric theory}

\author{Luc \textsc{Blanchet}}\email{luc.blanchet@iap.fr}
\affiliation{$\mathcal{G}\mathbb{R}\varepsilon{\mathbb{C}}\mathcal{O}$, 
	Institut d'Astrophysique de Paris,\\ UMR 7095, CNRS, Sorbonne Universit{\'e},\\
	98\textsuperscript{bis} boulevard Arago, 75014 Paris, France}
%\affiliation{Institut de Physique Th\'eorique, Universit\'e Paris-Saclay,\\
%CEA, CNRS, 91191 Gif-sur-Yvette, France}

\author{Lavinia \textsc{Heisenberg}} \email{laviniah@ethz.ch}
\affiliation{Institut f\"{u}r Theoretische Physik, ETH Z\"{u}rich,\\
Wolfgang-Pauli-Str. 27, 8093 Z\"{u}rich, Switzerland}
\affiliation{Institute for Theoretical Physics, Heidelberg University, Philosophenweg 16, 69120 Heidelberg, Germany}

\author{Fran\c{c}ois \textsc{Larrouturou}}\email{francois.larrouturou@desy.de}
\affiliation{Deutsches Elektronen-Synchrotron DESY, Notkestr. 85, 22607 Hamburg, Germany}

\date{\today}

\begin{abstract}

Motivated by a version of the ``Dipolar Dark Matter'' model, that aims at a relativistic completion of the phenomenology of MOND, we investigate the gravitational polarization mechanism in the canonical bimetric theory with an effective matter coupling. We explicitly show the fundamental obstacle why such theories cannot achieve a consistent gravitational polarization, and thus fail to recover the MONDian phenomenology at low energies.

\end{abstract}

\pacs{95.35.+d, 04.50.Kd}

\preprint{DESY-23-010}

\maketitle

\section{Introduction}
\label{sec:introduction}

Despite passing all current tests with flying colors (the last in date being the discovery of gravitational waves~\cite{GW150914,GW170817}), our theory of General Relativity (GR) suffers from some ``darkness'', \emph{i.e.}, yet ambiguous gravitational phenomena that can be explained either by invoking non-gravitational ingredients, or by ``beyond-GR'' completions~\cite{Heisenberg:2018vsk}. The paramount instance is that stars orbiting in periphery of galaxies do not move accordingly to GR predictions~\cite{Bosma, Rubin}, but instead behave as if the visible mass of the galaxy was only a small fraction of the total mass. The usual way to address this ``missing mass'' problem is to postulate the existence of some unknown non-relativistic form of matter, insensitive to electromagnetic forces (otherwise it would have already been seen), coined ``cold dark matter''~\cite{BHS05}. If this paradigm is a cornerstone of current physics, there is still no evidence of this new particle, despite decades of experimental efforts~\cite{Xenon20,Antares20,ATLAS21}.

Another way to explain the missing mass problem could be to consider that in some regime, the laws of gravity deviate from GR. This path has been initiated by the 40-years old seminal work of Milgrom~\cite{Milg1}, setting a very simple modification of Newton’s law that accounts for the motion of stars in galaxies and provides a natural explanation for the Tully-Fischer relation. Nevertheless, the MOND (for MOdified Newtonian Dynamics) approach is yet only phenomenological, despite great efforts to embed it within a viable fundamental theory (see \emph{e.g.}~\cite{Bek04,BGef07,DEFW14}, and~\cite{FamMcG12} for a comprehensive review on the MONDian approach). An alternative possibility to build such a relativistic completion is to exploit the dielectric analogy of MOND~\cite{B07mond}, which has led to several ``Dipolar Dark Matter'' (DDM) models~\cite{BL08,BL09,BB14,BH15a,BH15b,BH17}. In the present work, we focus on a subclass of such models~\cite{BH15a,BH15b,BH17} that are based on a bimetric extension of GR in which the two metrics are supposed to host opposite ``gravitational charges'' (see also~\cite{bimond1} for a different type of MONDian bimetric theory).

Thus the DDM phenomenology relies on a mechanism of gravitational polarization associated with opposite gravitational charges. It was indeed claimed in~\cite{BH15b}, that in the non-relativistic limit, using the canonical bimetric theory, the matter present in each sector feel forces with opposite signs, leading to the desired polarization. Nevertheless, as shown in the present work, due to a missing term in~\cite{BH15b}, unfortunately such polarisation mechanism cannot happen in this bimetric framework, which ruins this DDM model. However previously proposed DDM models~\cite{BL08,BL09} in standard GR are still viable. 

The plan of this paper is as follows. The DDM model used, which is an extension of the one studied in~\cite{BH15b,BH17}, is presented in Sec.~\ref{sec:DDM}. The proof of the impossibility of a gravitational polarization mechanism in this general set up is then exposed in Sec.~\ref{sec:proof}. Concluding remarks are presented in Sec.~\ref{sec:concl}.

\section{Model}\label{sec:DDM}

This work investigates an attempt for a Dipolar Dark Matter (DDM) model, generalizing the one used in previous works (see \emph{e.g.}~\cite{BH17}). Starting from the two dynamical metrics $\{g_{\mu\nu},f_{\mu\nu}\}$,\footnote{Through this work, we use a mostly positive signature for both metrics, denote space-time indices with greek letters and, once a proper 3+1 decomposition achieved, purely spatial indices with latin letters. For simplicity, we set $G = c= \hbar = 1$, unless otherwise specified.} our action reads
\begin{equation}\label{eq_Ltot}
\mathcal{S} = \int\!\!\dd^4x \,\mathcal{L}_\text{grav} +  \int\!\!\dd^4x \,\mathcal{L}_\text{mat} +  \int\!\! \dd^4x \,\mathcal{L}_\text{bar}\,,
\end{equation}
where $\mathcal{L}_\text{mat}$ describes the dark matter, coupled to both $g_{\mu\nu}$ and $f_{\mu\nu}$, and where $\mathcal{L}_\text{bar}$ is the usual baryonic matter, coupled to the metric $g_{\mu\nu}$. The gravitational sector is given by
\begin{equation}\label{eq_Lgrav}
\mathcal{L}_\text{grav} = \frac{M_g^2}{2}\sqrt{-g} \,R\left[g\right]+ \frac{M_f^2}{2}\sqrt{-f} \,R\left[f\right] + M_\text{eff}^2 m^2\,\mathcal{V}\left[g,f\right]\,,
\end{equation}
where $R$ denotes the usual Ricci scalar, $M_g$ and $M_f$ are the Planck scales associated with each metrics and $M_\text{eff}^2\,m^2$ is an unspecified scale. As for the interaction term $\mathcal{V}\left[g,f\right]$, we will naturally use the ghost-free potential function of de Rham, Gabadadze and Tolley~\cite{dRGT10,deRhamLR}, parametrized by five arbitrary constants $\{\alpha_n\}$ as
\begin{equation}\label{eq_VdRGT}
\mathcal{V}\left[g,f\right] =\sqrt{-g} \sum_{n=0}^4\alpha_n \,e_n\left(X\right)\,,
\end{equation}
where the square-root matrix $X^\mu_{\ \nu}$ is defined as $X^\mu_{\ \rho}X^\rho_{\ \nu} = g^{\mu\rho}f_{\rho\nu}$, together with its inverse $Y^\mu_{\ \nu}$, such that $Y^\mu_{\ \rho}Y^\rho_{\ \nu} = f^{\mu\rho}g_{\rho\nu}$, and $e_n(X)$ represent the elementary symmetric polynomials
\begin{equation}
	\begin{aligned}
		e_0(X) &= 
		1\,, \\ 
		e_1(X) &= 
		\bigl[X\bigr]\,, \\ 
		e_2(X) &=
		\frac{1}{2}\bigl(\bigl[X\bigr]^2-\bigl[X^2\bigr]\bigr)\,,\\ 
		e_3(X) &=
		\frac{1}{6}\bigl(\bigl[X\bigr]^3-3\bigl[X\bigr]\bigl[X^2\bigr]
		+2\bigl[X^3\bigr]\bigr)\,, \\ 
		e_4(X) &=
		\frac{1}{24}\bigl(\bigl[X\bigr]^4-6\bigl[X\bigr]^2\bigl[X^2\bigr]
		+3\bigl[X^2\bigr]^2 + 8\bigl[X\bigr]\bigl[X^3\bigr]-6\bigl[X^4\bigr]\bigr)\,,
	\end{aligned}
\end{equation}
where the brackets denote the trace operation. The interaction term~\eqref{eq_VdRGT} is a generalization of the one taken in previous works, corresponding to the choice $\alpha_n = \alpha^{4-n}\beta^n$, where $\alpha$ and $\beta$ enter the effective metric~\eqref{eq_effmetric}, see \emph{e.g.} Eq.~(2.8) of~\cite{BH15b}. The Lagrangian~\eqref{eq_Lgrav} with~\eqref{eq_VdRGT} is nothing but the gravitational sector of the canonical ghost-free bimetric theory of Hassan and Rosen~\cite{Hassan12a,Hassan12b}, where $\alpha_0$ (resp. $\alpha_4$) plays the role of a bare cosmological constant for the metric $g_{\mu\nu}$ (resp. $f_{\mu\nu}$). This gravitational sector is symmetric under the exchange of the metrics $g_{\mu\nu} \leftrightarrow f_{\mu\nu}$ together with the relabeling $\{ M_g,\alpha_n\} \leftrightarrow \{ M_f,\alpha_{4-n}\}$.

As for the matter sector of our model, in addition to the baryonic matter $\mathcal{L}_\text{bar}$ coupled to $g_{\mu\nu}$ in the usual way, we will consider the following Lagrangian
\begin{equation}\label{eq_Lmat}
\mathcal{L}_\text{mat} =  \mathcal{L}^g_\text{mat}[J_g^\mu; g_{\mu\nu}] + \mathcal{L}^f_\text{mat}[J_f^\mu; f_{\mu\nu}] + \mathcal{L}^\text{eff}_\text{mat}[A_\mu; g^\text{eff}_{\mu\nu}] + \mathcal{L}^\text{int}_\text{mat}[J_{*g}^{\mu}, J_{*f}^{\mu}, A_\mu]\,,
\end{equation}
that describes three matter fields: (i) particles coupled to $g_{\mu\nu}$ and described by a conserved\footnote{With the conserved current we can describe particles without interaction or a perfect fluid with zero temperature. Adding a dependence of the specific entropy we could also describe the case with finite temperature, which we will not consider in this work.} current $J_g^\mu$ (such that $\nabla^g_\mu J_g^\mu=0$); (ii) particles coupled to $f_{\mu\nu}$ and described by a conserved current $J_f^\mu$ (such that $\nabla^f_\mu J_f^\mu=0$); (iii) an Abelian $U(1)$ vector field $A_\mu$ coupled to the effective metric~\cite{dRHRa}
\begin{equation}\label{eq_effmetric}
g^\text{eff}_{\mu\nu}=\alpha^2 g_{\mu\nu} +2\alpha\beta
\,g_{\mu\rho}X^\rho_{\ \nu} +\beta^2 f_{\mu\nu}\,,
\end{equation}
where $\alpha$ and $\beta$ are arbitrary constants. The last term entering the matter sector~\eqref{eq_Lmat} is an interaction between the particles and the $U(1)$ field
\begin{equation}\label{eq_Lint}
\mathcal{L}^\text{int}_\text{mat} = \bigl(J_{*g}^{\mu} - J_{*f}^{\mu}\bigr) A_\mu\,.
\end{equation}
where $J_{*g}^\mu \equiv \sqrt{-g}\,J_g^\mu$ and $J_{*g}^\mu \equiv \sqrt{-f}\,J_f^\mu$ are metric independent matter currents.

Note that the interaction term~\eqref{eq_Lint} induces a supplementary, although indirect, coupling between the two metrics, and can thus reintroduce the ghost that was suppressed by the interaction term~\eqref{eq_VdRGT}~\cite{dRHRa,dRHRb}. Restricting the interaction term to the case $\alpha_n = \alpha^{4-n}\beta^n$ as was done previously, it has been shown that the ghost is not excited for energies beyond the decoupling limit of the theory~\cite{BH17}. In our more general framework~\eqref{eq_VdRGT}, it could be that the theory is plagued by a ghost at energies well below the decoupling limit. Nevertheless, as our aim is to prove that those theories are not well suited for a MONDian phenomenology, and as such a ghost would definitively not play a role in this low-energy analysis, we will consider the larger class of theories with~\eqref{eq_Lint} and arbitrary coefficients $\{\alpha_n\}$.

\section{Proof of the absence of polarization mechanism}\label{sec:proof}

Let us now turn to the study of the weak-field (non-relativistic) limit of the theory~\eqref{eq_Ltot}, investigated along the lines of~\cite{BH15b}. In order to reach a consistent weak-field limit, we assume that the Universe is empty in average, \emph{i.e.} that the matter only enters at perturbative order in the action. Moreover, seeking for a viable gravitational polarization mechanism requires to perturb both metrics around a common background metric, say $\bar{g}_{\mu\nu}$. Note that, as we will see later on, an empty background is not necessarily a flat one, as the interaction term will play the role of an effective cosmological constant. Thus $\bar{g}_{\mu\nu}$ is not the Minkowskian metric, but rather a de Sitter (or anti-de Sitter) one. We then perturb each metric as
\begin{equation}\label{eq_gfpert}
\begin{aligned}
g_{\mu\nu} &= (\bar{g}_{\mu\nu}+h_{\mu\nu})^2 
\equiv (\bar{g}_{\mu\rho}+h_{\mu\rho})\bar{g}^{\rho\lambda}(\bar{g}_{\lambda\nu}+h_{\lambda\nu})\,,\\ 
f_{\mu\nu} &=(\bar{g}_{\mu\nu}+\ell_{\mu\nu})^2
\equiv (\bar{g}_{\mu\rho}+\ell_{\mu\rho})\bar{g}^{\rho\lambda}(\bar{g}_{\lambda\nu}+\ell_{\lambda\nu})\,.
\end{aligned}
\end{equation}
Up to quadratic order in the metric perturbations, the Lagrangian becomes\footnote{We do not write here the zeroth order in perturbation, naturally irrelevant for our study. From now on, the barred quantities are associated with the background metric $\bar{g}_{\mu\nu}$, with which the indices are operated.}
\begin{equation}\label{eq_Lpert}
\begin{aligned}
\mathcal{L}^{(2)} = 
\sqrt{-\bar{g}} \, \biggl\{
&
- \Bigl(M_g^2 h^{\mu \nu} + M_f^2 \ell^{\mu\nu}\Bigr) \overline{G}_{\mu\nu}
- \Bigl(M_g^2\Lambda_g h^{\mu\nu} + M_f^2 \Lambda_f \ell^{\mu\nu}\Bigr)\bar{g}_{\mu\nu}\\
&
- M_g^2 h^{\mu\nu}\overline{\mathcal{E}}^{\rho\lambda}_{\mu\nu} h_{\rho\lambda}
- M_f^2 \ell^{\mu\nu}\overline{\mathcal{E}}^{\rho\lambda}_{\mu\nu}\ell_{\rho\lambda} 
+ h_{\mu\nu}\Bigl(T_\text{bar}^{\mu\nu}+T_g^{\mu\nu}\Bigr)
+ \ell_{\mu\nu} \,T_f^{\mu\nu}\\
&
+ \frac{M_\text{eff}^2\,m^2}{2}
\Bigl[ \Gamma_0 h^{\mu\nu}h^{\rho\lambda}
+ 2\Gamma_1h^{\mu\nu}\ell^{\rho\lambda}
+ \Gamma_2\ell^{\mu\nu}\ell^{\rho\lambda} \Bigr]\bigl(\bar{g}_{\mu\nu}\bar{g}_{\lambda\rho} - \bar{g}_{\mu\rho}\bar{g}_{\lambda\nu} \bigr)\biggr\}\,,
\end{aligned}
\end{equation}
where $\overline{G}_{\mu\nu}$ is the Einstein tensor associated to $\bar{g}_{\mu\nu}$, and $\overline{\mathcal{E}}^{\rho\sigma}_{\mu\nu}$ is the Lichnerowicz operator
\begin{equation}\label{eq_Lichne}
\begin{aligned}
-2\overline{\mathcal{E}}^{\rho\lambda}_{\mu\nu}h_{\rho\lambda} &=
\overline{\Box} \bigl(h_{\mu\nu} - \bar{g}_{\mu \nu} h \bigr)
+ \overline{\nabla}_{\mu}\!\overline{\nabla}_{\nu} h
- 2 \overline{\nabla}_{(\mu} H_{\nu)}
+ \bar{g}_{\mu \nu} \overline{\nabla}_{\rho} H^{\rho} \\
&
- 2 \overline{C}_{\mu\rho\lambda\nu} h^{\rho\lambda}
- \frac{2}{3} \,\overline{R} \,\Bigl(h_{\mu \nu} - \frac{1}{4} \bar{g}_{\mu \nu} h \Bigr)
- \overline{G}_{\rho (\mu} h^\rho_{\nu)}\,.
\end{aligned}
\end{equation}
Here $\overline{C}_{\mu\nu\rho\lambda}$ is the Weyl tensor, we denoted the trace as $h \equiv h^{\rho\sigma}\bar{g}_{\rho\sigma}$ and posed $H_\rho \equiv \overline{\nabla}^{\sigma}h_{\rho\sigma}$. The stress-energy tensors are defined as usual by
\begin{equation}\label{eq_Tmunu}
T^{\mu\nu}_\text{bar} = \frac{2}{\sqrt{-g}}\frac{\delta \mathcal{L}_\text{bar}}{\delta g_{\mu\nu}}\,,\qquad
T^{\mu\nu}_g = \frac{2}{\sqrt{-g}}\frac{\delta \mathcal{L}_\text{mat}}{\delta g_{\mu\nu}}\,,\qquad
T^{\mu\nu}_f = \frac{2}{\sqrt{-f}}\frac{\delta \mathcal{L}_\text{mat}}{\delta f_{\mu\nu}}\,.
\end{equation}
They are first-order quantities (as we required an empty Universe) that obey conservation laws with respect to the background metric at leading order, namely
\begin{equation}\label{consTmunu}
	\overline{\nabla}_\nu \Bigl(T^{\mu\nu}_\text{bar} + T^{\mu\nu}_g\Bigr) = \mathcal{O}(2)\,,\qquad \overline{\nabla}_\nu T^{\mu\nu}_f = \mathcal{O}(2)\,.
\end{equation}
Here we neglect the stress-energy tensor associated with the internal vector field $A_\mu$, which is coupled to the effective metric~\eqref{eq_effmetric} and is second order $\mathcal{O}(2)$ in perturbation, see~(3.5) in~\cite{BH15b}. Finally, we have introduced the effective cosmological constants
\begin{equation}\label{effLambda}
	\Lambda_g 
	= -\frac{M_\text{eff}^2\,m^2}{M_g^2}\big(\alpha_0+3\alpha_1+3\alpha_2+\alpha_3\big)\,,
	\qquad
	\Lambda_f 
	= -\frac{M_\text{eff}^2\,m^2}{M_f^2}\big(\alpha_1+3\alpha_2+3\alpha_3+\alpha_4\big)\,,
\end{equation}
as well as the combinations
\begin{equation}\label{comb}
\Gamma_0 \equiv \alpha_0+2\alpha_1+\alpha_2\,,\qquad
\Gamma_1  \equiv \alpha_1+2\alpha_2+\alpha_3\,,\qquad
\Gamma_2  \equiv \alpha_2+2\alpha_3+\alpha_4\,.
\end{equation}

The Lichnerowicz operator considered in Ref.~\cite{BH15b} had a missing term: Eq. (3.10) there missed the last term in~\eqref{eq_Lichne}, \emph{i.e.} $- \overline{G}^{}_{\rho (\mu} h^\rho_{\nu)}$, proportional to the Einstein tensor of the background. This term plays a major role when establishing the relation~\eqref{eq_div_Lichne} below, and in turn the Eq.~\eqref{eq_cond_hl}, instead of the relation (3.13) found in~\cite{BH15b}. Unfortunately, this invalidates the model~\cite{BH15b} for the proposed application to MONDian phenomenology.

At the linear level, the variation of the perturbed Lagrangian~\eqref{eq_Lpert} gives the field equations
\begin{equation}\label{eq_cond_Lambda}
\overline{G}_{\mu\nu} = - \Lambda_g\,\bar{g}_{\mu\nu} = - \Lambda_f \,\bar{g}_{\mu\nu}\,,
\end{equation}
which can be fulfilled only for a common cosmological constant in the background, $\Lambda_g = \Lambda_f$. This equality is nothing but the condition that the background metric $ \bar{g}_{\mu\nu}$ is indeed the same for both sectors. When translated in terms of the notation~\eqref{comb},
%coefficients $\alpha_n$, 
this condition gives a generalisation of Eq.(3.8) of~\cite{BH15b}:
\begin{equation}\label{eq_cond_alpha}
\frac{\Gamma_0+\Gamma_1}{M_g^2} = \frac{\Gamma_1+\Gamma_2}{M_f^2}\,.
\end{equation}

Turning now to the quadratic level, the Einstein field equations for the two metric perturbations $h_{\mu\nu}$ and $\ell_{\mu\nu}$ read
\begin{equation}\label{eq_eom_hl}
\begin{aligned}
& 
2 M_g^2\,\overline{\mathcal{E}}^{\mu\nu}_{\rho\lambda}h^{\rho\lambda} + M_\text{eff}^2\,m^2 \biggl[ \Gamma_0 \bigl(h^{\mu\nu}-h\,\bar{g}^{\mu\nu}\bigr)+\Gamma_1\bigl(\ell^{\mu\nu}-\ell\,\bar{g}^{\mu\nu}\bigr)\biggr] = T_\text{bar}^{\mu\nu} + T_g^{\mu\nu}\,,\\
& 
2 M_f^2\,\overline{\mathcal{E}}^{\mu\nu}_{\rho\lambda}\ell^{\rho\lambda} + M_\text{eff}^2\,m^2 \biggl[\Gamma_1 \bigl(h^{\mu\nu}-h\,\bar{g}^{\mu\nu}\bigr)+\Gamma_2\bigl(\ell^{\mu\nu}-\ell\,\bar{g}^{\mu\nu}\bigr)\biggr] = T_f^{\mu\nu}\,.
\end{aligned}
\end{equation}
As usual in bigravity, we observe that the mass eigenstates are different from the kinetic ones. Using the expression~\eqref{eq_Lichne} as well as the background condition~\eqref{eq_cond_Lambda} which says that $\Lambda \equiv \Lambda_g = \Lambda_f$, the divergence of the Lichnerowicz operator reads
\begin{equation}\label{eq_div_Lichne}
2\overline{\nabla}_\nu\left(\overline{\mathcal{E}}^{\mu\nu}_{\rho\lambda} h^{\rho\lambda}\right) =
2\overline{G}_{\rho\lambda} \overline{\nabla}^{[\mu}_{}h^{\rho]\lambda}_{}
+ \overline{\nabla}_\rho\Bigl(\overline{G}_{}^{\lambda[\mu}h^{\rho]}_{\lambda}\Bigr)
= \Lambda\,\overline{\nabla}_\rho \bigl[h^{\rho\mu} - h \,\bar{g}^{\rho\mu}\bigr]\,,
\end{equation}
together with the mirror relation obtained by replacing $h^{\rho\sigma}\rightarrow \ell^{\rho\sigma}$.
%naturally holds for $\overline{\nabla}_\nu\left(\overline{\mathcal{E}}^{\mu\nu}_{\rho\lambda} \ell^{\rho\lambda}\right)$. 
Therefore, using the conservation laws~\eqref{consTmunu} of the stress-energy tensors, we find that the divergence of both field equations~\eqref{eq_eom_hl} boil down to the same constraint, independent of the background relation~\eqref{eq_cond_alpha},
\begin{equation}
\Gamma_1 \overline{\nabla}_\rho \big[h^{\rho\mu} - \ell^{\rho\mu} - h\,\bar{g}^{\rho\mu} + \ell\,\bar{g}^{\rho\mu}\big] = 0\,.
\end{equation}
Imposing $\Gamma_1 = 0$, the two sectors would decorrelate at first order, as clear from Eq.~\eqref{eq_eom_hl}, and thus no gravitational polarization mechanism could be possible, by definition. Therefore, we are forced to impose
\begin{equation}\label{eq_cond_hl}
\overline{\nabla}_\rho \big[h^{\rho\mu} - \ell^{\rho\mu}\big] = \overline{\nabla}^\mu\big[h- \ell\big]\,.
\end{equation}

We now investigate the phenomenological consequences of this constraint in the non-relativistic limit $c\to\infty$. In this regime, we consider Newtonian-like systems, with sizes significantly smaller that the effects of a reasonable cosmological constant. We can thus discard the effect of the effective cosmological constant $\Lambda$, and take a Minkowskian ansatz for $\bar{g}_{\mu\nu}$, which allows a proper 3+1 decomposition of the components. On this flat background, we parametrize the metric perturbations by single Newtonian potentials $U_g$ and $U_f$ and two ``PPN'' constants $\gamma_g$ and $\gamma_f$ as
\begin{equation}
\begin{aligned}
& 
h^{00} = \frac{U_g}{c^2}+\mathcal{O}\left(\frac{1}{c^4}\right)\,,\qquad
h^{0i} = \mathcal{O}\left(\frac{1}{c^3}\right)\,,\qquad
h^{ij} = \gamma_g\,\frac{U_g}{c^2}\,\delta_{ij}+\mathcal{O}\left(\frac{1}{c^4}\right)\,,\\
&
\ell^{00} = \frac{U_f}{c^2}+\mathcal{O}\left(\frac{1}{c^4}\right)\,,\qquad
\ell^{0i} = \mathcal{O}\left(\frac{1}{c^3}\right)\,,\qquad
\ell^{ij} = \gamma_f\,\frac{U_f}{c^2}\,\delta_{ij}+\mathcal{O}\left(\frac{1}{c^4}\right)\,.
\end{aligned}
\end{equation}
The $ij$ components of the field equations~\eqref{eq_eom_hl} set $\gamma_g = \gamma_f = 1$, and the constraint equation~\eqref{eq_cond_hl} then gives
\begin{equation}\label{eq_cond_U}
\partial_i \big[U_g-U_f\big] = 0\,.
\end{equation}
However, the gravitational polarization mechanism discussed in~\cite{BH15b} required that the gradients of $U_g$ and $U_f$ have opposite signs: as the particles described by $J_g^\mu$ and $J_f^\mu$ feel the force respectively due to the potentials $U_g$ and $U_f$, they will be displaced in the opposite direction. The $U(1)$ field $A_\mu$ coupling both sectors would then create an effective polarization of the spacetime medium, reproducing the main feature of the model~\cite{B07mond}. It is thus clear that the constraint~\eqref{eq_cond_U} is incompatible with such a mechanism.

\section{Conclusion}\label{sec:concl}

This study has proven that, contrarily to what was claimed in the work~\cite{BH15b}, the framework based on canonical bimetric theory is unable to produce a consistent gravitational polarization mechanism. This shortage unfortunately spoils its capacity to account for an effective MONDian phenomenology, and thus, to be a proper relativistic completion of the MOND paradigm. Note that another DDM model, based on standard GR~\cite{BL08,BL09}, is still viable in this respect; this model has recently been simulated numerically on galactic scales~\cite{SFTD22}.
Nevertheless, there is still hope that a viable mechanism could be achieved in a ghost-free bigravity framework, as DDM was based on the Hassan-Rosen model, but other bimetric theories exist, for instance the newly proposed minimal bigravity~\cite{DFLMM20}, which has the advantage of having stable cosmological solutions as well as escaping the Higuchi ghost present in usual bigravity scenarii.

\acknowledgments

F.L. would like to thank the Institut d'Astrophysique de Paris for hosting him during both early and late stages of this project.

\bibliography{ListeRef_DDM_pol.bib}

\end{document}